\documentclass[aps,prb,twocolumn,floatfix,showpacs]{revtex4}
\usepackage{graphicx}
\usepackage{amsmath}
\usepackage{amssymb}
\usepackage{epsfig}
\newcommand\beq{\begin{equation}}
\newcommand\eeq{\end{equation}}
\newcommand\bear{\begin{eqnarray}}
\newcommand\eear{\end{eqnarray}}
\def\ep{{\epsilon}}
\def\om{{\omega}}
\def\he{{h_{e\!f\!\!f}}}
\def\omt{{\tilde{\omega}}}

\def\cg{{\cal{G}}}

\begin{document}
\title{Interplay between strong correlations and magnetic field in the \\symmetric
periodic Anderson model}
\author{Debabrata Parihari} 
\author{N.\ S.\ Vidhyadhiraja}
\affiliation{Theoretical Sciences Unit,\\Jawaharlal Nehru Centre for
Advanced Scientific Research,\\Jakkur, Bangalore 560064, India.}
\email{raja@jncasr.ac.in}
\author{David E.\ Logan}
\affiliation{Oxford University, Chemistry Department, Physical and Theoretical Chemistry Laboratory,\\South Parks Road,
Oxford OX1 3QZ, UK.}

\date{\today}

\begin{abstract}
\parbox{6in}
{
Magnetic field effects in Kondo insulators are studied theoretically, using a
local moment approach to the periodic Anderson model within the framework of 
dynamical mean-field theory. Our main focus is on 
field-induced changes in single-particle dynamics and the associated 
hybridization gap in the density of states.
Particular emphasis is given to the strongly correlated regime, where 
dynamics are found to exhibit universal scaling in terms of a 
field-dependent low energy coherence scale. Although the bare applied field is globally
uniform, the effective fields experienced by the conduction electrons
and the $f$-electrons differ because of correlation effects. A continuous insulator-metal transition is found to occur on increasing the applied field, closure of the hybridization
gap reflecting competition between Zeeman splitting and screening of the $f$-electron local
moments. For intermediate interaction strengths the hybridization gap depends non-linearly
on the applied field, while in strong coupling its field dependence is found to be linear.
For the classic Kondo insulator YbB$_{12}$,  good agreement is found upon 
direct comparison of the field evolution of the experimental transport gap
with the theoretical hybridization gap in the density of states.
}
\end{abstract}
\maketitle
\section{Introduction}
Kondo insulator materials such as SmB$_6$, YbB$_{12}$ and Ce$_3$Bi$_4$Pt$_3$
have been of sustained interest to experimentalists and theorists for
several decades~\cite{grewe,hewson,aeppli,fisk,takabatake,degiorgi,peter}. Interest 
in these system stems from their unusual properties, such as the small hybridization gap 
(of a few meV) and mixed valency, as well as the  transition to metallic behavior with
doping~\cite{dopedki}, application of pressure~\cite{presski} and magnetic field~\cite{magki}. The underlying qualitative origin of such rich behavior is the strong electronic correlation arising due to the localized and narrow $f$-orbitals of the rare earth atoms, which hybridize weakly with broad, essentially non-interacting conduction bands. 

A quantitative understanding of the dynamics and transport properties of these materials has not however been easy to achieve. 
A principal stumbling block in this regard has been the theoretical
treatment of localized and itinerant fermionic degrees of freedom on a comparable
footing. Much progress has been made in recent years with the advent of dynamical mean field theory (DMFT)~\cite{vollhard,jarrell,kotliar,gebhard}, within which generic lattice-fermion models such as the Hubbard or the periodic Anderson model 
have found approximate solutions, and quantitative agreement with experiments 
has also been obtained~\cite{raja_ki,kotliar}. Within DMFT~\cite{vollhard,jarrell,kotliar,gebhard}, a lattice fermion model 
is mapped onto an effective single-site correlated impurity which hybridizes
with a self-consistent conduction electron bath. Thus,
various techniques such as the numerical renormalization group, exact 
diagonalization, diagrammatic perturbation theory based approaches, quantum
Monte Carlo and many others that have, in the past, been developed to handle 
the many-body single-impurity problem, have now been adapted and modified for 
use within the DMFT framework~\cite{vollhard,jarrell,kotliar,gebhard}. One such recent technique is the local moment approach (LMA)~\cite{lmaintro,NLDSR,lmaff,lmanrg,victoria,raja_ki,rajahf,epjb}, which has been shown to be powerful not only in the context of single-impurity systems~\cite{lmaintro,lmaff,lmanrg,NLDSR}, but also 
for lattice-based heavy fermion systems~\cite{raja_ki,rajahf,victoria,epjb} when used in conjunction with DMFT. In this paper, we employ the LMA to understand the interplay of electronic correlations and an external magnetic field in Kondo insulator
materials.

      The generic model used to study Kondo insulator materials is the 
periodic Anderson model (PAM), which consists in physical terms of a correlated $f$-level 
in each unit cell hybridizing locally with a non-interacting conduction band. Magnetic field effects in these systems have been studied theoretically through the inclusion of a Zeeman
term in the PAM~\cite{saso,lee}. The observed insulator-metal transition~\cite{magki} has also been reproduced qualitatively in theoretical  calculations~\cite{saso}. However, a detailed 
understanding of the changes in single-particle dynamics and the associated hybridization gap has on the whole been lacking, and a quantitative description of the experimentally observed
field-induced behavior has not been achieved. We seek to bridge these gaps
in this paper by studying the PAM with a Zeeman term using LMA+DMFT, and with
particular emphasis on the strongly correlated (or strong coupling) regime. Our primary
focus is on the field-induced changes in the single-particle dynamics and the associated 
hybridization gap in the density of states. 

The outline of the paper is as follows: We begin in the next section with 
a brief description of the model (PAM), the DMFT framework and the LMA
technique for the PAM in the presence of a magnetic field.  In section 3 we present our theoretical results and their analysis. We also make comparison between  theory and experiments on the classic Kondo insulator material $YbB$$_{12}$. Brief conclusions are given in section 4. 

\section{Model and Formalism}
The Hamiltonian for the PAM is given in standard notation by
\bear
H = -\sum_{\langle ij\rangle,\sigma} t_{ij}c_{i\sigma}^{\dagger}
c_{j\sigma}^{\phantom{\dagger}}+ \sum_{i\sigma}
(\epsilon_{f}+\frac{U}{2} f_{i-\sigma}^{\dagger}
f_{i-\sigma}^{\phantom{\dagger}})f_{i\sigma}^{\dagger}
f_{i\sigma}^{\phantom{\dagger}}\nonumber\\ +
V\sum_{i\sigma}(f_{i\sigma}^{\dagger}c_{i\sigma}^{\phantom{\dagger}}+{\rm h.c}) + \ep_c
\sum_{i\sigma} c_{i\sigma}^{\dagger}c_{i\sigma}^{\phantom{\dagger}}
\eear
where the first term describes the kinetic energy of the noninteracting 
conduction ($c$) band due to nearest neighbour hopping $t_{ij}$.
The second term refers to the $f$-levels with site energies $\ep_{f}$ and on--site repulsion 
$U$, while the third term describes the $c/f$ hybridization via the local matrix element $V$.
The final term represents the $c$-electron orbital energy.
Within DMFT~\cite{vollhard,jarrell,kotliar,gebhard}, which is exact in the 
limit of infinite dimensions, the hopping term is scaled as 
$t_{ij} \propto t_{*}/\sqrt{Z_{c}}$, with coordination number
$Z_{c} \rightarrow \infty$.In this paper we consider mainly the hypercubic lattice, for which
the non-interacting density of states is a Gaussian~\cite{kotliar}
($\rho_0(\ep)= \exp(-\ep^2/t_*^2)/\sqrt{\pi}t_{*})$. We also consider specifically the 
(particle-hole) symmetric PAM, which is the traditional limit 
employed to study Kondo insulators~\cite{kotliar,jarrell_ki,raja_ki,victoria,
saso,lee,shim,sun, pruschke,faze}. 
For the symmetric PAM the conduction band is located symmetrically about the Fermi level 
({\it i.e.\ }$\ep_c=0$), while $\epsilon_{f} = - U/2$. This corresponds to
 half-filling of the $f$ and $c$-levels, {\it i.e.\ }$n_{f} = \sum_{\sigma}\langle f_{i\sigma}^{\dagger}
f_{i\sigma}^{\phantom{\dagger}}\rangle
 = 1$ and $n_{c}=\sum_{\sigma}\langle c_{i\sigma}^{\dagger}
c_{i\sigma}^{\phantom{\dagger}}\rangle = 1$ for all $U$.
In this case the system is an insulator for all interaction strengths $U$ (in 
the absence of a magnetic field), with a corresponding gap in the 
single-particle spectra.
The symmetric PAM in the absence of a magnetic field has been studied 
quite comprehensively  within the DMFT framework, see 
e.g.~\cite{kotliar,jarrell,victoria,raja_ki,shim,jarrell_ki,sun}. 

Within DMFT, the lattice fermion model maps onto an effective, correlated single-site 
impurity hybridizing self-consistently with a conduction electron 
bath~\cite{vollhard,jarrell,kotliar,gebhard}. The
self energy is thus spatially local,  {\it i.e.\ }momentum
independent.  Thus, the problem is simplified to a great extent. Nevertheless,
the problem remains non-trivial because the impurity model is as yet
unsolved for an arbitrary hybridization. As mentioned in the introduction,
the local moment approach~\cite{lmaintro} has been successful in describing
the single impurity Anderson model~\cite{lmaintro,lmaff,lmanrg,NLDSR} as well as in 
understanding the PAM within DMFT~\cite{raja_ki,rajahf,victoria,epjb}.
Here we extend the approach to encompass finite magnetic fields in the
symmetric PAM, enabling study of magnetic field effects in Kondo insulators.

The local moment approach begins with the symmetry broken mean field state
(unrestricted Hartree Fock, UHF). Dynamical self energy effects are then
built in through the inclusion of transverse spin fluctuations for each of the 
symmetry broken states. The most important idea underlying the LMA at $T=0$ is
that of symmetry restoration~\cite{lmaintro,NLDSR,victoria}: self-consistent restoration
of the broken symmetry inherent at pure mean field level, arising in physical terms
via dynamical tunneling between the locally degenerate mean-field states,  and 
in consequence ensuring correct recovery of the local Fermi liquid behavior that reflects adiabatic continuity in $U$ to the  non-interacting limit.
The reader is referred to our earlier papers~\cite{raja_ki,victoria} for full
details of the formalism and implementation of the zero-field LMA for the PAM. 

 In the presence of a global magnetic field, the degeneracy between the mean 
field solutions~\cite{lmaintro} (labeled as A and B corresponding to $+|\mu|$
and -$|\mu|$) found at the mean field level is lifted, and only one 
solution as determined by $sgn(h)$ remains~\cite{lmaff}. Here, we consider
explicitly $h>0$ for which only the `A' solution survives. 
For $h\neq 0$ the bare electronic energy levels, 
$\epsilon_{\alpha}$ for $\alpha = c$ and $f$, are of course split 
via the Zeeman effect. The modified energy levels
are given by $\ep_{\alpha\sigma} = \ep_{\alpha} - \sigma h_\alpha$ with
$h_\alpha=\tfrac{1}{2}g_\alpha\mu_B H$ (and $\sigma = \pm$ for $\uparrow/$$\downarrow$ spins), where $\mu_B$ is the Bohr magneton, $g_\alpha$ the
Lande g-factor and $B$ the magnetic field. Although  $g_{f} \neq g_{c}$ in general, 
for simplicity we set $g_{f} = g_{c}=g$ {\it i.e.\ }we consider the
application  of a uniform magnetic field $h \equiv h_c=h_f$ to both $c$ and $f$-levels. 
The local (site-diagonal) Green functions for the $c$- and $f$-levels
may be expressed using the Feenberg renormalized perturbation 
theory~\cite{feenberg,economou} as 
\bear
G_{A \sigma}^{c}(\omega;h) &=& [\omega^{+} + \sigma h -
\frac{V^{2}}{\omega^{+} + \sigma h - \tilde{\Sigma}_{fA\sigma}(\omega;h)}
\nonumber\\ && - S_{A\sigma}[G_{A\sigma}^{c}(\omega;h)]]^{-1} 
\label{eq:gc}\\
G_{A \sigma}^{f}(\omega;h) &=& [\omega^{+} + \sigma h - 
\tilde{\Sigma}_{fA\sigma}(\omega;h) -\nonumber\\ 
&&\frac{V^{2}}{\omega^{+} + \sigma h
- S_{A\sigma}[G_{A\sigma}^{c}(\omega;h)]}]^{-1}.
\label{eq:gf}
\eear
where $\omega^+ =\omega +isgn(\omega)0^{+}$. Here $ S_{A\sigma}$ is the Feenberg self energy, 
a functional solely of $G_{A\sigma}^{c}$, given by
\beq
S_{A\sigma}(\om;h)=\gamma_{A\sigma}(\om;h) - \frac{1}{{\rm{H}}
[\gamma_{A\sigma}(\om;h)]} \label{eq:sw}\\
\eeq
with
\beq
\gamma_{A\sigma}(\om;h)=\omega^{+} + \sigma h - \frac{V^{2}}{\omega^{+}
+ \sigma h - \tilde{\Sigma}_{fA\sigma}(\omega;h)}\label{gamma}
\eeq
and the Hilbert transform ${\rm H}[z]$ defined as
\beq
{\rm{H}}[z]=\int^\infty_{-\infty}\frac{\rho_0(\ep)\,d\ep}{z-\ep}\,
\label{eq:HT}
\eeq
such that $G^{c}_{A\sigma}(\om;h) = H[\gamma_{A\sigma}]$.

\begin{figure}
\begin{center}
\rotatebox{0}{
\includegraphics[scale=0.5]{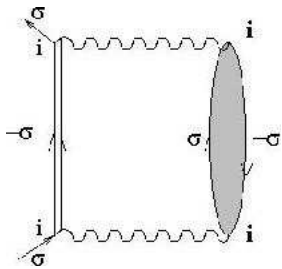}}
\end{center}
\caption{Principal contribution to the LMA $\{\Sigma_{f\sigma}(\omega)\}$.
Wavy lines denote U. See text for details.}
\label{self}
\end{figure}

Dropping the `A' subscript for clarity, the spin-summed Green functions are denoted by $G^{\alpha}(\omega;h) = \tfrac{1}{2}\sum_{\sigma}G_\sigma^{\alpha}(\omega;h)$ ($\alpha=c,f$), with corresponding spectral functions $D^{\alpha}(\omega;h) = -\frac{1}{\pi}sgn(\omega)
{\rm{Im}} G^{\alpha}(\omega;h)$. Within the LMA the $f$-electron self energies $\tilde{\Sigma}_{f\sigma}(\omega;h)$are familiarly separated  into a static mean-field
contribution plus a dynamical part $\Sigma_{f\sigma}(\omega;h)$~\cite{lmaintro,raja_ki,victoria}, 
\beq
\tilde{\Sigma}_{f\sigma}(\omega;h) = -\frac{\sigma}{2}U|\bar{\mu}(h)| +
\Sigma_{f\sigma}
(\omega;h)
\label{eq:self}
\eeq
where $|\bar{\mu}(h)|$ is the UHF local moment.
We approximate the dynamical part of the self energy by the usual non-perturbative 
class of  diagrams retained in practice by the LMA (as shown in figure~\ref{self}), which may be expressed mathematically at zero temperature~\cite{lmaintro,raja_ki,victoria}  as 
\beq
\Sigma_{f\sigma}(\omega;h) = U^{2}\int_{-\infty}^{\infty} \frac
{d\omega_{1}}{2\pi i}\;\cg_{-\sigma}(\omega-\omega_1;h)
\Pi^{-\sigma \sigma}(\omega_1;h).
\eeq
Here the host/medium Green function $\cg_\sigma$ (denoted by the double-lined 
propagator in figure~\ref{self}) is defined by
\beq
\cg_\sigma(\om;h)^{-1} = G^f_\sigma(\om;h)^{-1} + \Sigma_{f\sigma}(\om;h)\,.
\eeq
$\Pi^{-\sigma \sigma}(\omega;h)$ denotes the transverse spin 
polarization propagator (shown shaded in figure~\ref{self}), which in the random
phase approximation employed is expressed as 
 $\Pi^{+-} =\, ^{0}\!\Pi^{+-}/(1 - U\, ^{0}\!\Pi^{+-})$.
The bare polarization propagator $^{0}\!\Pi^{+-}(\omega;h)$ is expressed in 
terms of mean-field propagators~\cite{victoria,raja_ki} $\{g^{\alpha}_\sigma(\om;h)\}$ (in which only the static Fock contribution to the self-energies occurs).

For $h=0$ the symmetry restoration (SR) condition for the symmetric PAM
is given by~\cite{victoria,raja_ki} $\tilde{\Sigma}_{f\sigma}(\omega =0; h=0) =0$
(independent of spin $\sigma$, since by particle-hole symmetry 
$\tilde{\Sigma}_{f\sigma}(\omega ; h) =-\tilde{\Sigma}_{f-\sigma}(-\omega; h)$),
i.e. by
\beq
\Sigma_{f\uparrow}(\om=0;h=0)=U|\bar{\mu}|/2.
\label{eq:sr}
\eeq
Satisfaction of the SR condition ensures adiabatic continuity (in $U$) to the 
non-interacting limit~\cite{victoria,raja_ki} of the hybridization-gap insulator (the system
being in that sense a generalized Fermi liquid~\cite{victoria,raja_ki}); and 
as expected the system remains insulating, with a gap in the single-particle spectra,
for all interactions $U\geq 0$.

The practical implementation of the above procedure is carried out as follows~\cite{raja_ki,victoria,epjb}, in which the numerical procedure 
is simplified by keeping $x=\tfrac{1}{2}U|\mu|$ fixed, while varying
$U$ to satisfy the symmetry restoration condition.
We begin with an $h=0$ calculation for a given $x$. (i) The mean-field Green functions 
$g^{\alpha}_\sigma(\om;h)$ are obtained from equations (2-7) by retaining only the static part of the self energy. 
(ii) These Green functions are then used to construct the bare 
polarization bubble $^0\!\Pi^{+-}$. (iii) The DMFT iterative procedure starts with a
specific dynamical self energy, obtained either as a guess or from the previous
iteration, which is substituted into equations (2-7) to get the 
$c$- and $f$- Green 
functions. (iv) The host/medium Green function $\cg_\sigma(\om;h)$ is obtained 
through equation (9).  (v) The transverse spin polarization propagator 
$\Pi^{+-}$ is computed for a given $U$ and along with 
$\cg_\sigma(\om;h)$, substituted in equation (8) to determine the 
$\om=0$ self energy for a given $U$; (vi) The SR condition (equation (10)) is checked, and 
if found not to be satisfied, $U$ is varied and  step (v) is repeated until the SR condition is satisfied. (vii) Upon restoration of the broken spin
symmetry, the corresponding $U$ is obtained and equation (8) may then be used 
to get the full dynamical self energy at all frequencies. (viii) The new 
self energy is fedback in the first step of the DMFT procedure (step (iii))
and the iterations are continued until full self-consistency is achieved. 
For finite fields the same procedure is adopted except that SR is no longer
imposed~\cite{lmaff} (the $U$ found from SR at
$h=0$ is naturally used for all $h>0$).

\section{ Results and Discussion}
\label{results}

\begin{figure}[t]
\includegraphics[scale=0.32]{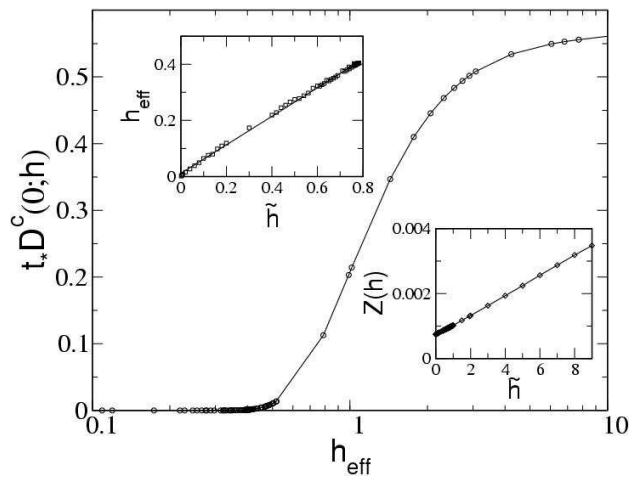}
\caption{ $c$-band spectrum at the Fermi level $D^c(0;h)$
in the strong coupling, universal regime as a function of $\he$,
showing a continuous insulator-metal transition
at $\he$$_{\!,c}\simeq 0.36$. Top inset: shows the linear dependence of
the effective field $\he$ on $\tilde{h}=h/\omega_{L}(0)$ ($\omega_{L}(0)=Z(0)V^{2}/t_{*}$).
Bottom inset: linear increase of the quasiparticle weight $Z(h)$ with increasing 
$\tilde{h}$.
}
\label{fig:dc0}
\end{figure}

Before discussing the interplay between interactions
and magnetic fields in the symmetric PAM, we consider briefly
two limiting cases. (i) First the non-interacting limit, $U=0$,
with $h\neq 0$; and secondly (ii) the interacting problem $U>0$ 
in the absence of a field.
In the former case, the non-interacting $c$-spectrum for spin-$\sigma$ is given by 
\begin{displaymath}
d^c_{0\sigma}(\om;h)=
\rho_0\left(\om +\sigma h - \frac{V^2}{\om+\sigma h}\right)\,.
\end{displaymath}
The spectral band edges are given by
\beq
\om +\sigma h - \frac{V^2}{\om+\sigma h} = \pm W
\eeq
where $2W$ is the full band-width of the non-interacting density of states $\rho_0(\omega)$.
It is easy to see from this that for $h=0$ there is a hybridization gap at the Fermi level (denoted by $\Delta_{0}(0)$), which 
decreases with increasing field $h$ and eventually closes at a
field value that is half of the zero field spectral gap, i.e.\
\beq
\Delta_{0}(h)=\Delta_{0}(0) -2h.
\label{eq:nigap}
\eeq
Thus in the non-interacting limit, application of a magnetic 
field leads to an insulator-metal transition simply because of the rigid
crossing of the up- and down-spin bands. We add here that the
non-interacting Gaussian density of states characteristic of the hypercubic
lattice (as considered explicitly below) is of course unbounded and as such
does not possess `hard' band edges. The field-induced insulator-metal transition
in this case is thus strictly a crossover, although in practice 
the transition is `sharp' as one would expect (see e.g.\ figure~\ref{fig:dc0}).

\begin{figure}
\includegraphics[scale=0.32]{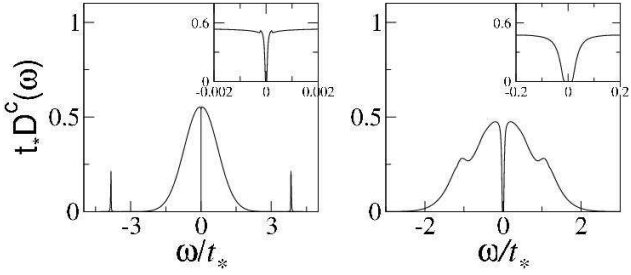}
\caption{LMA conduction electron spectra for $h=0$, 
in strong coupling (left panel,  $U/t_{*} = 7.1$)
and weaker coupling (right panel, $U/t_{*}= 1.2$), with $V^2=0.2t_*^2$. Insets 
show the low frequency region of the respective main panels.
}
\label{lma_h0}
\end{figure}

In the second limit, of finite interactions but zero field, 
the ground state remains gapped for all interaction strengths~\cite{victoria},
the hybridization gap $\Delta$ decreasing continuously with increasing
$U$ from its non-interacting limit $\Delta_{0}(0)$.
In the strong coupling, Kondo lattice regime of the model, 
universal scaling occurs~\cite{raja_ki,victoria,shim,pruschke,faze}
in terms of an exponentially small low energy scale $\om_L=ZV^2/t_* (\equiv \frac{1}{2}\Delta)$;
$Z$ being the quasiparticle weight or mass renormalization factor, given by $Z=(1-\partial\Sigma_{f\sigma}(\omega)/\partial
\om|_{\om=0})^{-1}$.
The Green functions and their associated spectra depend solely on 
$\omt=\om/\om_L $ in the universal scaling 
regime~\cite{raja_ki,victoria}. Representative  results for the zero-field density of states 
are shown in figure~\ref{lma_h0}; where the main panels show the conduction 
band density 
of states, $t_{*}D^c(\om)$ (solid lines), as a function of frequency, $\om/t_*$.
The right panels represent intermediate coupling ($U/t_{*}= 1.2$), while the
left panels are for strong coupling ($U/t_{*}=7.1$). The insets show a close-up of
the low frequency spectra where the hybridization gap at the Fermi level
($\om=0$) is evident at both weak and strong coupling. 

Now we consider both interactions and the field.  In the strong coupling regime,
in parallel to the $h=0$ limit, 
we expect universality to persist 
in terms of a field-dependent low energy scale, $\omega_{L} \equiv \omega_{L}(h)$.
To derive explicitly the universal scaling form in the limit of low-frequencies 
(i.e.\ close to the Fermi level), we perform a simple low-frequency `quasiparticle
expansion' of the self energy, retaining only its real part
($\Sigma^R_{f\sigma}$) to leading order in $\omega$; i.e.\
\beq
\Sigma^R_{f\sigma}(\om;h)=\Sigma^R_{f\sigma}(0;h) - \left(\frac{1}{Z(h)}-1\right)\om\,
\label{eq:lin}
\eeq
where $Z(h) =(1-\partial\Sigma^{R}_{f\sigma}(\omega;h)/\partial
\om|_{\om=0})^{-1}$ is the field-dependent quasiparticle weight (independent
of $\sigma$ since $\Sigma^{R}_{f\sigma}(\omega ; h) =-\Sigma^{R}_{f-\sigma}(-\omega; h)$ by particle-hole symmetry).

Substituting equation (13) into equations~(\ref{eq:gc}-\ref{eq:self}), we find
that the associated spectral functions are just renormalized versions of 
their non-interacting counterparts, being given by
\bear
D^c_\sigma(\om;h)&\stackrel{\om\rightarrow 0}{\rightarrow}&
\rho_0\left(-\frac{1}{\omt + \sigma \he}\right)
\label{univgc} \\
D^f_\sigma(\om;h)&\stackrel{\om\rightarrow 0}{\rightarrow}&\frac{t_*^2}{V^2}
\frac{1}{(\omt + \sigma \he)^2}~
D^c_\sigma(\om;h) \label{univgf} 
\eear
where $\omt=\om/\om_{L}(h)$, and the low-energy scale $\omega_{L}(h) = Z(h)V^2/t_{*}$
is thus defined (in direct parallel to the $h=0$ limit).
In obtaining equations~(\ref{univgc},\ref{univgf}) we have explicitly considered
the strong coupling scaling regime, of 
finite $\tilde{\omega} =\omega/\omega_{L}(h)$ and $h/\omega_{L}(h)$ in the formal
limit where the low-energy scale $\omega_{L} \rightarrow 0$ (so that `bare' factors of $\omega \equiv \tilde{\omega}\omega_{L}$ or $h$ are thus neglected).
$\he$ in equations~(\ref{univgc},\ref{univgf}) is given by
\beq
\he=
\left[h-\sigma \tilde{\Sigma}^R_{f\sigma}(0;h)  
\right]~\frac{t_*}{V^2}\,
\label{eq:heff}
\eeq
(being independent of $\sigma$, by symmetry);
or equivalently, using the symmetry restoration condition 
$\tilde{\Sigma}^{R}_{f\sigma}(0;0)=0$, by
\beq
\he=\left[h+\sigma\left( 
\tilde{\Sigma}^R_{f\sigma}(0;0)
-\tilde{\Sigma}^R_{f\sigma}(0;h) 
\right)\right]\, \frac{t_{*}}{V^{2}}.
\label{eq:heffsc}
\eeq
In physical terms, $\he$ represents a dimensionless effective field, and its primary field-dependence arises from that of the interaction self-energy (the `bare' factor of $h$
in equations~(\ref{eq:heff}) or (\ref{eq:heffsc}) can of course be dropped in the strict scaling limit, although we retain it for clarity). In fact a leading order Taylor expansion of 
equation~(\ref{eq:heff}) or (\ref{eq:heffsc}) gives $\he = c \tilde{h}$, where $\tilde{h} = h/\omega_{L}(0)$ is the field rescaled in terms of the $h=0$ low energy scale $\omega_{L}(h=0) = Z(0)V^{2}/t_{*}$, and $c=Z(0)/\tilde{Z}$ ($\sim{\cal{O}}(1)$) with ${\tilde{Z}}=
[1-\sigma (\partial\Sigma^{R}_{f\sigma}(0;h)/\partial h)_{h= 0}]^{-1}$ thus defined.
From this simple consideration we anticipate that $\he$ is just a rescaled version of the bare magnetic field itself,
and is on the order of $\tilde{h} = h/\omega_{L}(0)$ (as confirmed explicitly below,
see figure~\ref{fig:dc0}).
\begin{figure}[t]
\includegraphics[scale=0.32]{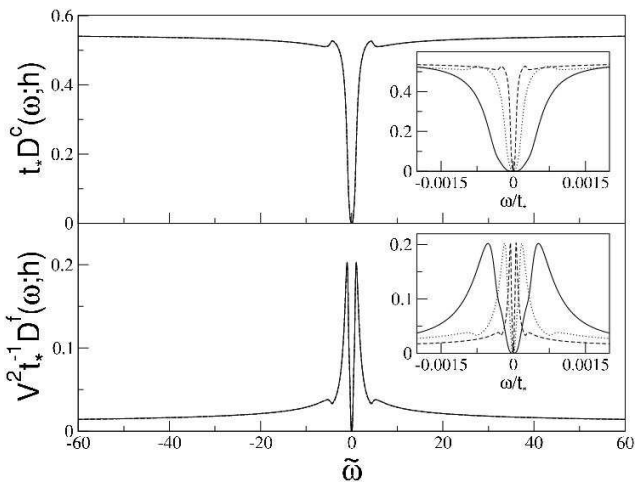}
\caption{ LMA spectra for the $c$-electrons (top panel,
$t_{*}D^{c}(\omega;h)$) and $f$-electrons (bottom panel,
$V^{2}t^{-1}_{*}D^{f}(\omega;h)$), for three parameter sets,
each corresponding to a fixed $\he=0.27$:
 $Ut_{*}/V^2=35, h/t_{*}=2.4\times10^{-5}$(dashed), $Ut_{*}/V^2=30, h/t_{*}=7.4\times10^{-5}$(dotted) 
 and $Ut_{*}/V^2=25, h/t_{*}=2.3\times10^{-4}$(solid).
The insets show the spectra 
{\it vs.\ }`bare' frequency, $\om/t_*$.
Main panel: The same spectra plotted {\it vs.\ }$\omt =\om /\omega_{L}(h)$
collapse to a common scaling form.
\label{fig:lwform}
}
\end{figure}

Equations~(\ref{univgc}) and (\ref{univgf}) 
show that in strong coupling, the spectra $D^{c}(\omega;h)$ and $V^{2}D^{f}(\omega;h)$ should be universal functions of $\tilde{\omega} =\omega/\omega_{L}(h)$,
for a fixed $\he$.
Thus, if distinct sets of model parameters in the strong coupling
regime correspond to the same $\he$, the 
spectra $D^c$ and $V^2D^f$ should collapse to the same scaling form
as a function of $\omt$, \emph{independently} of the bare parameters $U/t_{*}$ and
$V/t_{*}$. That this is is so is illustrated in figure
~\ref{fig:lwform}, where the top panel shows the full LMA $c$-electron spectra $t_{*}D^{c}$ for the hypercubic lattice, and the bottom panel the corresponding $f$-electron specta $(V^{2}/t_{*})D^{f}$. Three sets of spectra are shown, with
parameters $Ut_{*}/V^2=35$ and $h/t_{*}=2.4\times10^{-5}$(dashed), $Ut_{*}/V^2=30$ and 
$h/t_{*}=7.4\times10^{-5}$ (dotted) and $Ut_{*}/V^2=25$ with $h/t_{*}=2.3\times10^{-4}$(solid);
in each case, $\he$ ($=0.27$) is the same. The insets to the figure show that the spectra 
as a function of the `bare' frequency $\om/t_*$ are distinct.
However when plotted {\it vs.\ }$\omt$ (as shown in the main 
panels), they are indeed seen to collapse to a single universal form.

\begin{figure}[t]
\includegraphics[scale=0.32]{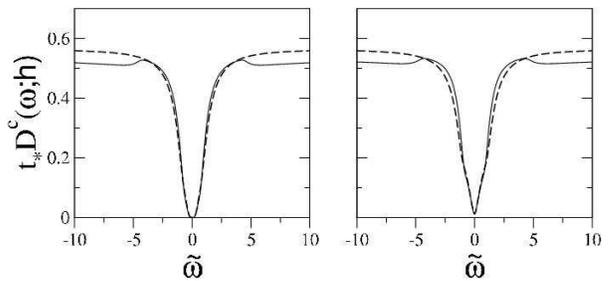}
\caption{Full LMA $c$-electron scaling spectra (solid lines) are 
compared to the limiting quasiparticle form equation~(\ref{univgc})
(dashed lines). The left panel shows the scaling spectra 
for $\he=0.27$ (where the system is an insulator) and the right panel for 
$\he=0.51$ (metallic).
}
\label{fig:ac}
\end{figure}

The quasiparticle forms in equations~(\ref{univgc},\ref{univgf}) embody local Fermi liquid behavior and adiabatic continuity to the non-interacting limit. They give explicitly the leading low-frequency asymptotic behavior of the scaling spectra that must be satisfied by any `full' theory. Direct comparison between the quasiparticle forms and the full LMA scaling spectra is shown in figure~\ref{fig:ac} for the $c$-electron spectra, and for two values of the effective field $\he$ (one corresponding to a case where the system remains insulating, the 
other for a higher field where the system is metallic, as discussed below). It is indeed clear from the figure that the LMA correctly recovers the limiting quasiparticle form in the vicinity of the Fermi level. In physical terms it is also worth noting that the low frequency quasiparticle  spectra are essentially those for the non-interacting limit, but with the local fields for $f$- and $c$-electrons replaced by $(V^2/t_{*})\he$ and zero respectively.
Hence, although the bare applied field is globally uniform, 
the effective local fields experienced by the $c$ and $f$ electrons are different because
of correlation effects. 

 We turn now to the transition with increasing field
from an insulating state characterised by a spectral gap straddling 
the Fermi level, to a metal with a finite density of states at $\omega =0$.
Equations~\ref{univgc} and~\ref{univgf} may be used to obtain an estimate of
the spectral band edges in strong coupling, and hence the gap as a function 
of the field. The band edges are given by 
\beq
\frac{1}{\omt+ \sigma\he} = \pm W/t_*
\eeq
with $2W$ the band-width of the non-interacting spectrum.
From this the field-dependent gap in $D^{c}(\om;h)$ or
$D^{f}(\om;h)$ follows as
\beq
\Delta (h)=2\left(1 - \he\frac{W}{t_*}\right)~\frac{Z(h)V^2}{W}\,.
\label{eq:gap_h}
\eeq
This in turn implies an insulator to  metal transition at a
critical effective field 
$\he$$_{\!,c}$ that is on the order of unity ($\sim t_*/W$).
The main panel in figure~\ref{fig:dc0} shows the variation of the full LMA 
density of states at the Fermi level, $D^{c}(\omega =0;h)$ (calculated explicitly for 
$U/t_{*} = 6.1, V^{2}/t_{*}^{2}=0.2$), as a function of $\he$ on a log scale. 
Although the calculations are for a hypercubic lattice, with a strictly soft 
gap in its zero-field spectrum, the insulator-metal transition is seen in practice 
to be sharp; occurring at a critical $\he$$_{\!,c} \simeq 0.36$ that is indeed
on the order of unity (we identify the critical field in practice from
$t_{*}D^{c}(0;h_{c}) \sim 10^{-3}$). On further increase of the field, 
$D^{c}(0;h)$ is seen from the figure to rise continuously, towards
the high-field value of $1/\sqrt{\pi}t_*$ which is just the non-interacting
dos value at the Fermi level. 

  The top inset to figure~\ref{fig:dc0} shows the dependence of the effective
field $\he$ (equations~(\ref{eq:heff},\ref{eq:heffsc})) on the scaled external field
$\tilde{h}=h/\omega_{L}(0)$ (with $\omega_{L}(0) =Z(0)V^2/t_{*}$). 
$\he$ is seen to be linear in $\tilde{h}$ (which behavior extends
over a wide $\tilde{h}$ interval) and, as anticipated above, is of the same
order as it: $\he \simeq \tilde{h}/2$ as evident from the figure.
The lower inset to the figure also  shows the $\tilde{h}$-dependence of the
quasiparticle weight $Z(h)$. It too is seen to increase linearly with 
field, implying a lowering of effective mass with an increase in field; and
which behavior is consistent with a similar finding for the single impurity Anderson model~\cite{lmaff}.

\begin{figure}[h]
\includegraphics[scale=0.32]{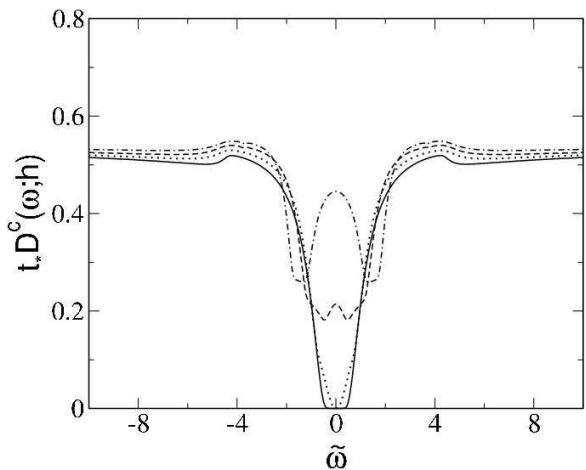}
\caption{ Universal $c$-electron spectra from the LMA are shown as a function of 
$\omt =\omega/\omega_{L}$ for various fields: $\he=0$ (solid), $\he=0.38$ (dotted), $\he=1$ (dashed) and $\he=5$ (dot-dashed). The closure of the insulating gap
with increasing field is evident.
}
\label{fig:dcw}
\end{figure}

The field-dependence of the full density of states is illustrated in
figure~\ref{fig:dcw}, where we plot the (universal, strong coupling) 
conduction band density of states $D^{c}(\om;h)$, as a function of 
the $\tilde{\om}$, for various $\he$. The solid curve $\he=0$ represents 
the insulating ground state,  while the dotted curve is for $\he=0.38$, which
is just above the insulator-metal transition, so the gap 
has closed. The remaining curves are for
$\he=1$ and $\he=5$, showing metallic densities of states 
characterised by a finite spectral density $D^c(0;h)$ at the Fermi level.

In the non-interacting limit, $U=0$, the  spectral gap closes 
linearly with the applied field as in equation~\ref{eq:nigap},
and the essential mechanism for the insulator-metal transition is
obvious: Zeeman splitting moves the up- and down-spin bands rigidly, 
resulting in their crossing at a critical field,
$h_{c0}/\Delta_{0}(0)=\tfrac{1}{2}$. This simple picture is naturally
modified in the presence of correlations, $U>0$, where two essentially competing
effects are operative. First, the tendency of the system to lower its energy
by uniform (`ferromagnetic') spin polarization of the $c$- and $f$-electrons,
i.e.\ the Zeeman effect, which alone operates in the non-interacting limit.
However for $h=0$ in the presence of interactions, lattice-coherent Kondo singlet
formation occurs, driven by local \emph{anti}ferromagnetic spin correlations between
the $c$- and $f$-electrons. In the presence of both interactions \emph{and} a field,
Zeeman splitting thus in effect competes with local moment screening; and the
field-dependence of the spectral gap is not a priori obvious.

\begin{figure}[h]
\includegraphics[scale=0.32]{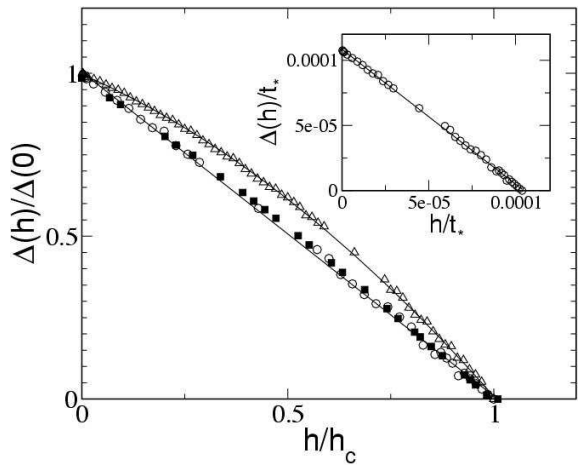}
\caption{The field-dependent spectral gap scaled by the zero field
gap, $\Delta (h)/\Delta (0)$,  {\it vs.\ }$h/h_c$, for
$U/t_{*}=1.2,V^2/t_{*}^{2}=0.2$ (triangles), $U/t_{*}=5.1,V^2/t_{*}^{2}=0.2$ (squares) and 
$U/t_{*}=6.1,V^2/t_{*}^{2}=0.2$ (circles). (The lines are best fits to the points.) Inset: the 
gap in units of $t_*$  \emph{vs.} the bare field $h/t_*$.
}
\label{fig:gap}
\end{figure}
LMA results for the spectral
gap are shown in figure~\ref{fig:gap} where the field dependent gap scaled by the
zero field gap, $\Delta (h)/\Delta (0)$, is plotted {\it vs.\ }$h/h_c$ 
for various interaction strengths. For intermediate coupling ($U/t_{*}=1.2,V^2/t_{*}^{2}=0.2$ (triangles)), the gap is seen to close non-linearly in the field and
is best fit by a quadratic form.
In the strong coupling regime by contrast 
(squares, $U/t_{*}=5.1,V^2/t_{*}^{2}=0.2$ and circles, $U/t_{*}=6.1,V^2/t_{*}^{2}=0.2$), linear 
behavior $\Delta (h)=\Delta (0)(1-h/h_c)$
is obtained, similar to the non-interacting limit. In this case however,
when $\Delta (h)$ is plotted directly {\it vs.\ }the bare field $h/t_{*}$ as
shown in the inset of figure~\ref{fig:gap}, the functional form
obtained is $\Delta (h)=\Delta (0)-h$; showing that the field
$h_c$ required to close the gap in strong coupling satisfies
$h_{c}/\Delta (0) = 1$, i.e.\
twice that required in the non-interacting limit, where 
$h_{c0}/\Delta_0(0)=\tfrac{1}{2}$. This result is physically
natural, in view of the effective competition between Zeeman splitting and
local moment screening discussed above.

Finally, we would like to make a comparison of our theoretical results 
to experiment.  For the classic Kondo insulator $YbB_{12}$, the field
dependence of the transport gap has been determined
from low-temperature resistivity measurements~\cite{magki} (the leading
low-$T$ behavior of the resistivity being $\rho(T) \propto \exp (-\Delta_{tr}/T)$
with $\Delta_{tr} \equiv \Delta_{tr}(h)$ the transport/activation gap).
Since the transport gap is known theoretically~\cite{raja_ki} to be proportional
to the spectral gap ($\Delta(0) \simeq 2\Delta_{tr}(0)$~\cite{raja_ki}),
comparison to experiment may be made.
\begin{figure}[h]
\includegraphics[scale=0.32]{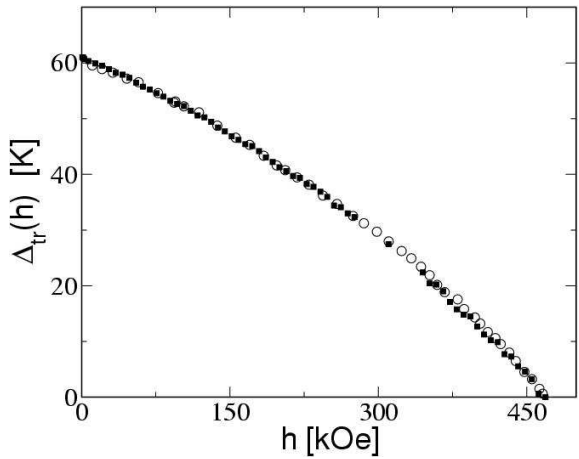}
\caption{Comparison of the experimental~\cite{magki} transport
gap in $YbB_{12}$ (opencircles) to the theoretical gap
(filled squares), obtained as discussed in text. 
The theory is seen to describe well the functional
form of the transport gap.
}
\label{fig:expt}
\end{figure}
In our earlier work~\cite{raja_ki} where we compared zero-field transport
properties of $YbB_{12}$ to theoretical results from the LMA,
we concluded that $YbB_{12}$ belongs to the intermediate coupling
regime (and as such lies outside the universal scaling regime). 
This is corroborated by the field-dependence of 
the transport gap, experimental results for which~\cite{magki} are shown
as open circles in figure~\ref{fig:expt}.
The dependence of $\Delta_{tr}(h)$ on the field $h$
is clearly non-linear, which behavior we have found above to be characteristic
of the intermediate coupling regime. To make comparison to experiment in
this regime, specific model parameters must of course be specified, and
here we choose $U/t_{*}=1.2, V^{2}/t_{*}^{2} = 0.2$ (the essential results
are quite insensitive to these particular values).
The filled squares in figure~\ref{fig:expt} show the field dependence 
of the resultant theoretical spectral gap, compared directly to experiment
with a simple multiplicative scaling of the $x$ and $y$ axes. The functional 
form of the theoretical gap is seen to be almost identical to that found 
experimentally, thus yielding good agreement between theory and experiment. 
Further, since the experimental $\Delta_{tr}(0) \simeq 60K$ (figure~\ref{fig:expt})
then the spectral gap $\Delta(0) \simeq 2 \Delta_{tr}(0) \simeq 120 K$; and for
the bare parameters considered we find $\Delta (0) = 0.026 t_{*}$. This in turn
yields the estimate $t_*\simeq 0.4eV$, which is physically realistic and compatible with
transfer integral values found through a band structure 
calculation~\cite{saso_band}.

\section{Conclusion}
The interplay between electronic correlations and an externally applied
magnetic field in Kondo insulators has been considered in this paper.
The symmetric periodic Anderson model, with a Zeeman term to account for 
the external magnetic field, has been studied within the
dynamical mean field framework using a local moment approach. In the strong coupling Kondo lattice regime of the model, the local $c$- and $f$-electron spectral functions are found to exhibit universal scaling, being functions solely of $\omega/\omega_{L}$ (with 
$\omega_{L}(h)=Z(h)V^{2}/t_{*}$ the characteristic low-energy scale) for a 
given effective field $\he$. Although the externally applied field is 
globally uniform, the effective local field experienced by the $c$- and $f$-electrons  differs 
because of correlation effects.  The zero-field spectral gap characteristic 
of  Kondo insulators is found to close  continuously, 
 leading to a continuous insulator-metal transition at a critical applied field 
$h_{c}$. Field induced closure of
the insulating gap is not simply a rigid band-crossing affair, but involves 
competition between local moment screening (reflecting correlation effects)
and Zeeman spin-polarization.
In the intermediate coupling regime the gap is found to close non-linearly with field, while 
in the strong coupling regime it closes linearly.
Comparison of the theoretical gap with the transport gap measured in the intermediate coupling material $YbB_{12}$ yields good agreement, providing support to the scenario presented 
for the field-induced gap closure. 

\begin{acknowledgments}
The authors DP and NSV would like to thank CSIR, India and JNCASR, India
while DEL would like to thank EPSRC, UK for supporting this research.
\end{acknowledgments}

\end{document}